# Quantum Nonlocality: Not Eliminated by the Heisenberg Picture

Ruth E. Kastner[†]

December 13, 2010

ABSTRACT. It is argued that the Heisenberg picture of standard quantum mechanics does not save Einstein locality as claimed in Deutsch and Hayden (2000). In particular, the EPR-type correlations that DH obtain by comparing two qubits in a local manner are shown to exist before that comparison. In view of this result, the local comparison argument would appear to ineffective in supporting their locality claim.

1. Introduction and background.

In a well-known paper, Deutsch and Hayden [1] argue that Einstein locality can be retained in quantum theory when processes are viewed in the Heisenberg picture instead of in the traditional Schrodinger picture. DH state that what they intend to preserve is Einstein's criterion that "the real factual situation of the system $S_2$ is independent of what is done with the system $S_1$, which is spatially separated from the former" ([2], p.85)). DH apparently take Einstein locality to be satisfied if the mathematical objects taken as representing each subsystem are factorizable, and if information about the setting of a distant measuring device can only be transmitted through classical (sublight) channels. But are these really sufficient conditions for Einstein locality? It is argued in this note that the factorizability of the operator expressions associated with the subsystems in the Heisenberg picture is not sufficient to preserve locality.[1] In particular, the EPR-type correlations that DH obtain by comparing two qubits in a local manner are shown to exist before the comparison.

First, let us review some background. In the standard Schrödinger picture, systems are labeled by time-dependent states such as

$|\psi(t)\rangle = \mathbf{U} |\psi(0)\rangle$,

$\mathbf{U}(t-t_0) = \exp[-i\mathbf{H}(t-t_0)/\hbar\,]$

(where $\mathbf{H}$ is the Hamiltonian operator and $\mathbf{U}(t-t_0)$ is the time evolution operator; operators are in bold face), while operators representing observables are time-independent. In the Heisenberg picture, one retains the initial quantum state $|\psi(0)\rangle$ and allows the time-dependence to attach to the operators instead, so a generic Schrödinger operator $\mathbf{X}$ becomes **X(t)**. Correspondence between the two pictures for computation of observable quantities such as probabilities is obtained by defining

$$\mathbf{X}(t) = \mathbf{U}^\dagger \mathbf{X} \mathbf{U}.$$

---

[†]rkastner@umd.edu; UMCP Foundations of Physics Group

[1] I should hasten to add that I do not think 'saving locality' is necessarily a worthwhile goal for an interpretation of quantum theory; indeed, it is most likely that what quantum theory is telling is that reality *is* nonlocal. So the failure of the Heisenberg picture to save locality is not regarded here as disappointing. On the contrary, it reinforces the idea that 'saving locality' should not be considered a criterion for the success or failure of any particular interpretation.

Observable quantities such as expectation values are the same in both pictures, since

$$\langle \mathbf{X}(t) \rangle = \langle \psi(t) | \mathbf{X} | \psi(t) \rangle_{Sch} = \langle \psi(0) | \mathbf{U}^\dagger \mathbf{X} \mathbf{U} | \psi(0) \rangle = \langle \psi(0) | \mathbf{X}(t) | \psi(0) \rangle_{Heis} \quad (1)$$

Thus, the dynamics is the same in both pictures; in the Heisenberg picture, we are just viewing the evolution of the system in a 'rotating frame' with respect to the Hilbert space (see Figure 1).

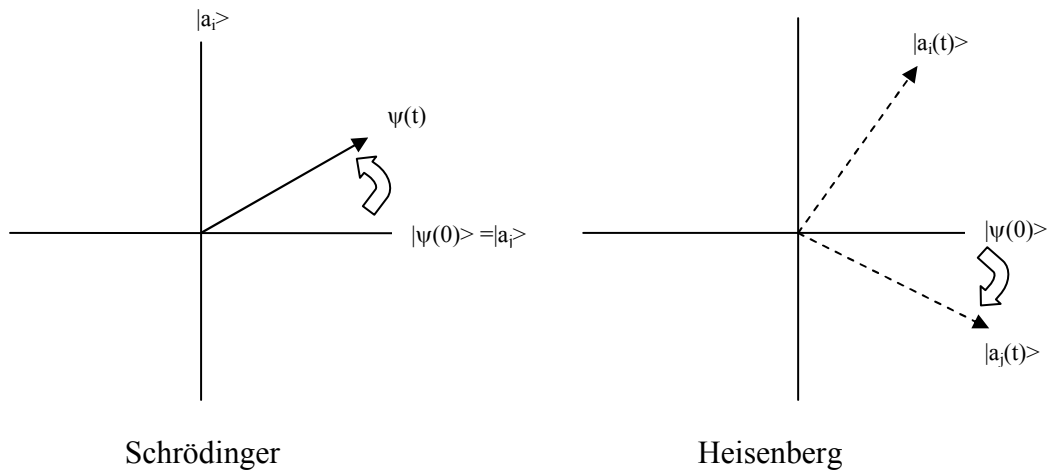

Schrödinger                                   Heisenberg

Figure 1. In the Schrödinger picture, the state vector rotates counterclockwise in the Hilbert space with increasing time index *t*. In the Heisenberg picture, the state vector remains stationary while the observable basis vectors rotate clockwise with increasing time *t*. The angular separation between the state vector and the basis vectors at a given time is the same in both pictures. Thus all observable quantities, which depend only on that relative separation, are independent of which picture is used.

2. Qubits and nonlocality in the Heisenberg picture

It should first be noted that DH adapt the traditional Heisenberg picture by introducing a unitary transformation by which they can define a 'standard state' devoid of any information concerning the initial physical state of the quantum systems. This procedure is particularly convenient for quantum computing, in which one can carry out operations on a network of qubits defined with respect to a standard computing basis state, usually defined as the "0" state for all qubits. Using this technique, DH present a version of the EPR-Bohm experiment in the context of a quantum computation with 4 qubits $Q_i$, i = [1,4] (see Figure 2). $Q_1$ and $Q_4$ are measuring ancillas which store outcomes, and $Q_2$ and $Q_3$ are measured systems which are placed into a Bell-type state through an inverse Bell gate. The latter consists of a Hadamard gate applied to $Q_3$ followed by a controlled-not (CNOT) gate applied to both qubits, with $Q_2$ as the target qubit (the qubit whose value is toggled, see below). The action of the Hadamard gate on an individual qubit is represented by

$$H = \frac{1}{\sqrt{2}} \begin{bmatrix} 1 & 1 \\ 1 & -1 \end{bmatrix} \quad (2)$$

The CNOT gate toggles the value of the ancilla (measuring) qubit if the measured (system) qubit has the value 1. Thus, it acts on the composite Hilbert space of $Q_2$ and $Q_3$ as

$$CN = \begin{bmatrix} Z- & Z+ \\ Z+ & Z- \end{bmatrix}, \quad (3)$$

where $Z^\pm$ are the projection operators for the states $|1\rangle$ and $|0\rangle$ respectively.

The basis with which CN is defined above is:

$$|1,1\rangle = \begin{pmatrix} 1 \\ 0 \\ 0 \\ 0 \end{pmatrix}; \quad |1,0\rangle = \begin{pmatrix} 0 \\ 1 \\ 0 \\ 0 \end{pmatrix}; \quad |0,1\rangle = \begin{pmatrix} 0 \\ 0 \\ 1 \\ 0 \end{pmatrix}; \quad |0,0\rangle = \begin{pmatrix} 0 \\ 0 \\ 0 \\ 1 \end{pmatrix} \quad (4)$$

Thus, for example, the action of CN on the composite state $|0,1\rangle$ (the second index being the measured qubit's value) is

$$CN \begin{pmatrix} 0 \\ 0 \\ 1 \\ 0 \end{pmatrix} = \begin{bmatrix} 0 & 0 & 1 & 0 \\ 0 & 1 & 0 & 0 \\ 1 & 0 & 0 & 0 \\ 0 & 0 & 0 & 1 \end{bmatrix} \begin{pmatrix} 0 \\ 0 \\ 1 \\ 0 \end{pmatrix} = \begin{pmatrix} 1 \\ 0 \\ 0 \\ 0 \end{pmatrix} = |1,1\rangle. \quad (5)$$

If all qubits are in the initial state $|0000\rangle$, after the inverse Bell gate, $Q_2$ and $Q_3$ are in the entangled state

$$|\Psi\rangle = (1/\sqrt{2})\ [\ |11\rangle - |00\rangle\ ] \quad (6)$$

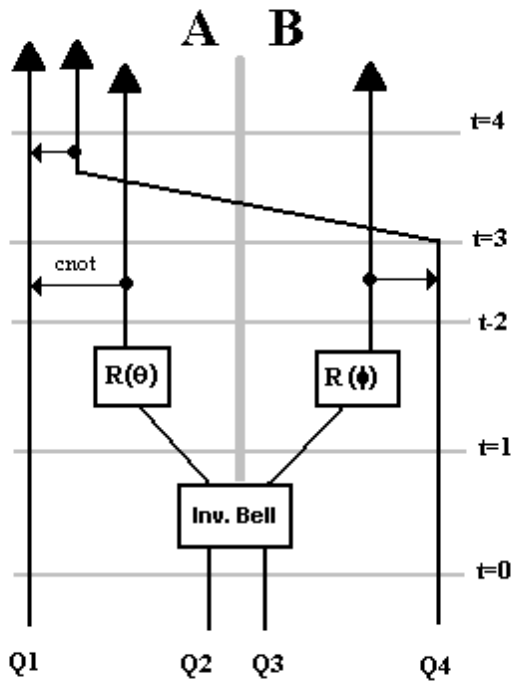

Figure 2: The EPR-Bohm experiment for qubits
as presented by Deutsch and Hayden (2000).

Now, DH proceed to calculate the probability that, for arbitary analyzer settings $\theta$ and $\phi$, the results were different (i.e., that $Q_1$ ends up in the state $|1>$), by using the information stored in $Q_1$ after the comparison measurement.[2] In doing so, they seem to imply that this is how the correlations are established, and that the latter are therefore local; this is not the case, as is shown below. While it is indeed true that measuring $Q_2$ before the comparison measurement cannot reveal any information about $\phi$ (and vice versa), this is just the no-signaling theorem, which is not under dispute (this fact has been noted previously by Timpson [3]). The question concerns the status of the correlations between $Q_2$ and $Q_3$. We can just as well calculate the probability that $Q_2$ and $Q_3$ are both, say, in the state $|1>$ after being subject to arbitary rotations $\theta$ and $\phi$. That is, nothing stops us from calculating and predicting this quantity even if it is performed by two different observers; we can just assume that $Q_2$ and $Q_3$ are subject to interaction Hamiltonians resulting in the appropriate change of basis, and calculate the probability that joint measurements of $Q_2$ and of $Q_3$ would reflect the result $|1,1\rangle$. This probability is given by[3]

$$\langle z_2(2)\, z_3(2) \rangle = \langle \tfrac{1}{2}(1+q_{z2})\, \tfrac{1}{2}(1+q_{z3}) \rangle$$

$$= \tfrac{1}{4} + \tfrac{1}{4} \langle q_{z2} + q_{z3} + q_{z2}\, q_{z3} \rangle. \qquad (7)$$

---

[2] The expectation values are calculated with respect to the computation basis state $|0,0>$. There appears to be a sign error in DH's result (28), since the probability that the outcomes were different for the state $|\Psi>$ is $\sin^2(\theta-\phi)/2$. Specifically, note that $<z_1(4)> = ½ - ½ <q_{1z}(3)q_{4z}(3)> =$
$½ - ½ (\cos\theta\cos\phi + \sin\theta\sin\phi) = ½ - ½ (\cos(\theta-\phi)) = \sin^2(\theta-\phi)/2$, not $\cos^2(\theta-\phi)/2$ as stated in DH eqn. (28) (arxiv version) or $\sin^2(\theta+\phi)/2$ (published version). (The expectation values for $I \otimes \sigma_z$ and $\sigma_z \otimes \sigma_z$ differ by a sign.)

Using DH's equation (25) for the associated observables at time 2 (suppressing the time index and the operators for $Q_1$ and $Q_4$):

$$q_{z2} = \sin\theta\, \sigma_{y2} \otimes \sigma_{x3} - \cos\theta\, \sigma_{z2} \otimes \sigma_{x3} \qquad (8a,b)$$

$$q_{z3} = \cos\phi\, I_2 \otimes \sigma_{x3} + \sin\phi\, \sigma_{x2} \otimes \sigma_{y3},$$

we find that the linear terms in (7) vanish. The remaining term yields (see footnote 1)

$$\langle q_{z2}\, q_{z3} \rangle = (\cos\theta\cos\phi + \sin\theta\sin\phi) = \cos(\theta-\phi) \qquad (9)$$

So that

$$\langle z_2(2)\, z_3(2) \rangle = \tfrac{1}{4} + \tfrac{1}{4}\cos(\theta-\phi) = \tfrac{1}{2}[\tfrac{1}{2} + \tfrac{1}{2}\cos(\theta-\phi)]$$

$$= \tfrac{1}{2}\cos^2(\theta-\phi)/2 \qquad (10)$$

which is, of course, the standard quantum mechanical probability for finding both qubits in the state $|1\rangle$ (or both spin-1/2 particles in the state "up") for the specified state (recall that this is not the standard singlet state). For example, if $\theta=\phi$, the probability for finding both systems "up" is $\tfrac{1}{2}$.

Thus, the correlations which DH argue are established through the local comparison measurement are clearly established at t=2, which undercuts DH's claim following their eqn. (28), applying to t=4, that "This ... is a familiar result, but in the course of calculating it in the Heisenberg picture, we have discovered exactly how the information about [the distant setting] reached $Q_1$: it was carried there in the qubit $Q_4$ as it travelled from B to A." The relevant "familiar result"—e.g., eqn. (10) above—was already established before this subluminal voyage.

Perhaps DH would argue that, in the Everettian picture, it doesn't matter that distant correlations are established, since all outcomes occur and therefore "the conditions of Bell's Theorem do not apply" (Rubin [4] (2001), p. 319). In that case, a particular kind of Everettian story (e.g., dynamically active labels matching up correlated distant outcomes, as proposed by Rubin [4]) would need to be invoked. Without that account, the local propagation story between the ancillas plays no role in establishing locality because it does not eliminate the nonlocal correlations existing before the comparison.

---

[3] The quantities denoted by subscripted letters are operators; we leave off the caret for simplicity. We also suppress the time index (t=2) in the calculation.

3. Conclusion

It has been argued, contra Deutsch and Hayden (2000), that Einstein locality is not saved by recourse to the Heisenberg picture, since distant correlations still demonstrably exist which violate Bell's inequality. These nonlocal correlations can be experimentally confirmed (whether they can be directly observed or not at a particular time). Thus factorization of the Heisenberg operators associated with individual subsystems is not enough to save locality. An appeal to a particular type of Everettian picture must be made, in which locality is preserved through information contained and transferred in a dynamical labeling process.

Acknowledgements.

The author is grateful for valuable comments by three anonymous reviewers.